\documentclass[10pt,conference,letterpaper]{IEEEtran}

\usepackage{times,amsmath}
\usepackage{amsfonts}
\usepackage{amssymb,bm}
\usepackage{amsthm}
\usepackage{algorithm}
\usepackage{algpseudocode}
\usepackage{cite}
\usepackage{graphicx}
\usepackage{epstopdf}
\usepackage[caption=false,font=footnotesize]{subfig}
\usepackage{fixltx2e}

\begin{document}
%
\title{Sum Rate Maximization for MU-MISO with Partial CSIT using Joint Multicasting and Broadcasting}

\author{
\IEEEauthorblockN{Hamdi Joudeh\IEEEauthorrefmark{1} and Bruno Clerckx\IEEEauthorrefmark{1}\IEEEauthorrefmark{2}}
\fontsize{9}{9}\upshape
\IEEEauthorrefmark{1} Department of Electrical and Electronic Engineering, Imperial College London, United Kingdom \\
\IEEEauthorrefmark{2} School of Electrical Engineering, Korea University, Seoul, Korea \\
\fontsize{9}{9}\selectfont\ttfamily\upshape
Email: \{hamdi.joudeh10, b.clerckx\}@imperial.ac.uk
}

\maketitle

\begin{abstract}
In this paper, we consider a MU-MISO system where users have highly accurate Channel State Information (CSI), while the Base Station (BS) has partial CSI
consisting of an imperfect channel estimate and statistical knowledge of the CSI error.
With the objective of maximizing the Average Sum Rate (ASR) subject to a power constraint, a special transmission scheme is considered where the BS  transmits a common symbol in a multicast fashion, in addition to the conventional private symbols.
This scheme is termed Joint Multicasting and Broadcasting (JMB).
The ASR problem is transformed into an augmented Average Weighted Sum Mean Square Error (AWSMSE) problem which is solved using Alternating Optimization (AO).
The enhanced rate performance accompanied with the incorporation of the multicast part is demonstrated through simulations.
\end{abstract}


\IEEEpeerreviewmaketitle

\section{Introduction}
\newcounter{Proposition_Counter} 
\newcounter{Lemma_Counter} 
\newcounter{Remark_Counter} 
\newcounter{Assumption_Counter}
The availability of accurate Channel State Information at the Transmitter (CSIT) is crucial for Downlink (DL) Multi-User (MU) multi-antenna wireless transmission.
This stems from the necessity to deal with the interference through preprocessing at the transmitter side, as the receivers are distributed \cite{Clerckx2013}.
While the ability to provide highly accurate and up-to-date CSIT remains questionable, considerable effort has been devoted to improving the performance in the presence of CSIT uncertainties.
Recent information theoretic developments focusing on the Multiple Input Single Output (MISO) Broadcast Channel (BC) suggest that multicast assisted transmission (where common symbols which are decodable by all users are transmitted alongside the conventional private symbols) can be used to enhance the performance in the infinite Signal to Noise Ratio (SNR) regime \cite{Yang2013,Hao2013}.
This paper focuses on the particular case where linear precoding is employed to transmit one common symbol in addition to private symbols in each channel use.
The simultaneous utilization of the MU-MISO medium as a Multicast Channel (MC) \cite{Jindal2006a} and a BC \cite{Viswanath2003} is termed Joint Multicasting and Broadcasting (JMB).

For a CSIT error that decays with increased SNR, JMB was shown to boost the achievable sum Degrees of Freedom (DoF) \cite{Yang2013,Hao2013}.
However, these results are intrinsically focused on the asymptotic SNR regime, where the DoF analysis is
most meaningful.
Therefore, trivial choices of linear precoders are deemed sufficient given that they achieve the aspired sum DoF.
For example, naive Zero Forcing Beamforming (ZF-BF) is used for the private symbols, while the multicast precoder is not optimized.
However, this is not the case at finite SNR where more involved performance metrics are considered, e.g. the Sum Rate (SR).
Works that consider the BC and MC separately under simpler CSI assumptions (i.e. perfect CSI) suggest that the choice of precoders can significantly influence the performance \cite{Christensen2008,Sidiropoulos2006}.
However, the instantaneous SR cannot be considered as a design metric at the BS due to the CSI uncertainty.
Alternatively, the Average\footnote{The term "Average" is used to denote the expectation w.r.t the CSIT error.} SR (ASR) is considered as an overall performance metric.
ASR maximization problems are tackled by extending the approach in \cite{Christensen2008,Shi2011}, i.e. transforming them into augmented Average Weighted Sum Mean Square Error (AWSMSE) problems which are solved using Alternating Optimization (AO) \cite{Bashar2014,Razaviyayn2013a}.

\emph{Contribution and Organization}:
In this work, we employ JMB transmission to optimize the SR performance for a MU-MISO system with partial CSIT.
To the best of our knowledge, this has not been considered in literature.
Due to the stochastic nature of the CSIT uncertainty, precoders are designed such that the ASR is maximized.
The problem is transformed into an equivalent augmented AWSMSE problem, solved using an AO algorithm which converges to a stationary point.
Moreover, we demonstrate the benefits of incorporating the common symbol. In particular, it is shown that the asymptotic DoF gains translate into SR gains in the finitely high SNR regime.
On the other hand, JMB reduces to conventional MU transmission when the common symbol is not needed, e.g. at low SNRs.

The rest of the paper is organized as follows: the system model and problem formulation are described in Section \ref{Section_System_Model}.
In Section \ref{Section_ASR_AWSMSE_Optimization}, the equivalent augmented AWSMSE problem is introduced.
An AO algorithm that solves the AWSMSE problem is proposed in Section \ref{Section_AO}. Simulation results are given in Section \ref{Section_Numerical_Results}, and Section \ref{Section_conclusion} concludes the paper.

\emph{Notation}: Boldface uppercase letters denote matrices, boldface lowercase letters denote column vectors and standard letters denote scalars. The superscrips $(\cdot)^{T}$ and $(\cdot)^{H}$ denote transpose and conjugate-transpose (Hermitian) operators, respectively. $\mathrm{tr}(\cdot)$ and $\|\cdot\|$ are the trace and Euclidian norm operators, respectively. Finally, $\mathrm{E}_{x}\{\cdot\}$ denotes the expectation w.r.t the random variable $x$.
%
\section{System Model and Problem Formulation}
\label{Section_System_Model}
We consider a Base Station (BS) equipped with $N_t$ antennas serving $K$ ($K \leq N_{t}$) single-antenna users.
The BS operates in a JMB fashion transmitting $K$ private symbols, each intended solely for one user, in addition to a common symbol that is decodable by all users.
The vector of complex data symbols is given as $\mathbf{s} \triangleq [s_{\mathrm{c}},s_{1},\ldots,s_{K}]^{T} \in\mathbb{C}^{K+1}$ where
$s_{\mathrm{c}}$ is the common symbol, $s_{i}$ is the $i$th user private symbol, $i\in\mathcal{K}$ and $\mathcal{K} \triangleq  \{1,\ldots,K\}$.
Entries of $\mathbf{s}$ have zero-means, unity powers and are mutually uncorrelated such that $\mathrm{E}\{\mathbf{s}\mathbf{s}^{H}\}=\mathbf{I}$.
$\mathbf{s}$ is linearly precoded into the transmit vector $\mathbf{x} \in\mathbb{C}^{N_{t}} $ given as
\vspace{-2.0mm}
\begin{equation}\label{Eq_x}
  \mathbf{x} = \mathbf{p}_{\mathrm{c}}s_{\mathrm{c}} + \sum_{i=1}^{K}\mathbf{p}_{i}s_{i}
\end{equation}
%
where $\mathbf{p}_{\mathrm{c}}\in\mathbb{C}^{N_{t}}$ and $\mathbf{p}_{i}\in\mathbb{C}^{N_{t}}$ correspond to the precoders for the common symbol and the $i$th private symbol respectively, from which $\mathbf{P} \triangleq  \big[\mathbf{p}_{\mathrm{c}},\mathbf{p}_{1},\ldots,\mathbf{p}_{K}\big]$ is composed.
The total transmit power at the BS is denoted as $P_{t}$, from which the transmit power constraint is given as $\mathrm{E}\{\mathbf{x}^{H}\mathbf{x}\} = \mathrm{tr}\big(\mathbf{P}\mathbf{P}^{H}\big)\leq P_{t}$.
For the $k$th user, the received signal is given as
\vspace{-1.0mm}
\begin{equation}
\label{Eq_yk}
  y_{k} = \mathbf{h}_{k}^{H}\mathbf{x}+n_{k}
\end{equation}
where $\mathbf{h}_{k} \in \mathbb{C}^{N_{t}}$ is the narrow-band channel impulse response vector between the BS and the $k$th user, from which the composite channel is defined as $\mathbf{H} \triangleq [\mathbf{h}_{1},\ldots,\mathbf{h}_{K}]$.
$n_{k} \thicksim \mathcal{CN} ( 0 , \sigma^{2}_{n_{k}} )$ is the AWGN at the $k$th receiver with variance $\sigma_{n_{k}}^{2}$. Throughout the paper, it is assumed that noise variances are equal across all users i.e. $\sigma_{n_{k}}^{2}=\sigma_{n}^{2}, \ \forall k \in \mathcal{K}$.
\subsection{CSIT Uncertainty}
Each of the $K$ links exhibits independent fading, and remains almost constant over a frame of symbols, enabling users to estimate their channel vectors with high accuracy.
On the other hand, CSIT experiences uncertainty arising from limited feedback, delays or mismatches.
$\mathbf{H}$ is written as a sum of the transmitter-side channel estimate
$\widehat{\mathbf{H}}\triangleq[\widehat{\mathbf{h}}_{1},\ldots,\widehat{\mathbf{h}}_{K}]$
and the channel estimation error
$\widetilde{\mathbf{H}}\triangleq [\widetilde{\mathbf{h}}_{1},\ldots,\widetilde{\mathbf{h}}_{K}]$,
such that
%
$ \mathbf{H}= \widehat{\mathbf{H}} + \widetilde{\mathbf{H}}$.
%
The CSIT consists of $\widehat{\mathbf{H}}$, in addition to some statistical knowledge of $\widetilde{\mathbf{H}}$.
Particularly, the BS knows the probability distribution of the actual channel given the available estimate, i.e.
$f_{\mathbf{H}|\widehat{\mathbf{H}}}(\mathbf{H})=f_{\widetilde{\mathbf{H}}}(\mathbf{H}-\widehat{\mathbf{H}})$.
%
\subsection{MSE, MMSE and Rate}
\label{Subsection_MSE_MMSE_Rate}
The $k$th user obtains an estimate of the common symbol by applying a scalar equalizer $g_{\mathrm{c},k}(\mathbf{h}_{k})$ to
\eqref{Eq_yk} such that $\widehat{s}_{\mathrm{c},k}=g_{\mathrm{c},k}(\mathbf{h}_{k}) y_{k}$.
Assuming that the common symbol is successfully decoded by all users, the common symbol's receive signal part is reconstructed and cancelled from $y_{k}$. This improves the detectability of $s_{k}$, which is then estimated by applying $g_{k}(\mathbf{h}_{k})$ such that
$\widehat{s}_{k}=g_{k}(\mathbf{h}_{k}) (y_{k}-\mathbf{h}_{k}^{H}\mathbf{p}_{\mathrm{c}}s_{\mathrm{c},k})$.
The notations $g_{\mathrm{c},k}(\mathbf{h}_{k})$ and $g_{k}(\mathbf{h}_{k})$ are used to emphasise the dependencies on the actual channel, as each user is assumed to have perfect knowledge of its own channel vector.
$(\mathbf{h}_{k})$ is omitted for brevity unless special emphasis is necessary. This is used with other variables that depend on the actual channel.
For the $k$th user, the MSEs  defined as $\varepsilon_{\mathrm{c},k} \triangleq \mathrm{E}_{\mathbf{s},n_{k}}\{|\widehat{s}_{\mathrm{c},k} - s_{\mathrm{c}}|^{2}\}$ and $\varepsilon_{k} \triangleq \mathrm{E}_{\mathbf{s},n_{k}}\{|\widehat{s}_{k} - s_{k}|^{2}\}$ are given as
\vspace{-2.0mm}
\begin{subequations}
\label{Eq_MSE}
\begin{align}
  \label{Eq_MSE_c_k}
  \varepsilon_{\mathrm{c},k}(\mathbf{h}_{k}) =& \ |g_{\mathrm{c},k}|^{2} T_{\mathrm{c},k} -2\Re \big\{g_{\mathrm{c},k}\mathbf{h}_{k}^{H}\mathbf{p}_{\mathrm{c}}\big\}+1 \\
  \label{Eq_MSE_k}
  \varepsilon_{k}(\mathbf{h}_{k}) =& \ |g_{k}|^{2} T_{k}-2\Re \big\{g_{k}\mathbf{h}_{k}^{H}\mathbf{p}_{k}\big\}+1
\end{align}
\end{subequations}
where $T_{\mathrm{c},k} = |\mathbf{p}_{\mathrm{c}}^{H}\mathbf{h}_{k}|^{2}+T_{k}$
and $T_{k} = \sum_{i=1}^{K}|\mathbf{p}_{i}^{H}\mathbf{h}_{k}|^{2}+\sigma_{n}^{2}$.
The optimum $g_{\mathrm{c},k}$ and $g_{k}$ are obtained by setting
$\frac{\partial \varepsilon_{\mathrm{c},k} }{\partial g_{\mathrm{c},k}}$ and $\frac{\partial \varepsilon_{k} }{\partial g_{k}}$ to zeros, yielding the well known MMSE equalizers:
\vspace{-1.0mm}
\begin{equation}
 \label{Eq_g_MMSE}
  g_{\mathrm{c},k}^{\mathrm{MMSE}} \! (\mathbf{h}_{k}) \! = \! \mathbf{p}_{\mathrm{c}}^{H}\mathbf{h}_{k} T_{\mathrm{c},k}^{-1}
  \ \text{and} \
  g_{k}^{\mathrm{MMSE}} \! (\mathbf{h}_{k}) \! = \! \mathbf{p}_{k}^{H}\mathbf{h}_{k}T_{k}^{-1}.
\end{equation}
Substituting (\ref{Eq_g_MMSE}) into (\ref{Eq_MSE}), the $k$th user's MMSEs are given as
\vspace{-1.0mm}
\begin{equation}
  \label{Eq_MMSE}
  \varepsilon_{\mathrm{c},k}^{\mathrm{MMSE}}(\mathbf{h}_{k}) = T_{\mathrm{c},k}^{-1} E_{\mathrm{c},k}
  \quad \text{and} \quad
  \varepsilon_{k}^{\mathrm{MMSE}}(\mathbf{h}_{k}) = T_{k}^{-1}E_{k}
\end{equation}
where $E_{\mathrm{c},k} =  T_{\mathrm{c},k} - |\mathbf{p}_{\mathrm{c}}^{H}\mathbf{h}_{k}|^{2} = T_{k}$ and
$E_{k} =  T_{k} - |\mathbf{p}_{k}^{H}\mathbf{h}_{k}|^{2}$.

The MMSE and the Signal to Interference plus Noise Ratio (SINR) are related such that $\gamma_{\mathrm{c},k}={(1-\varepsilon_{\mathrm{c},k}^{\mathrm{MMSE}})}/{\varepsilon_{\mathrm{c},k}^{\mathrm{MMSE}}}$ and $\gamma_{k}={(1-\varepsilon_{k}^{\mathrm{MMSE}})}/{\varepsilon_{k}^{\mathrm{MMSE}}}$, where $\gamma_{\mathrm{c},k}$ and $\gamma_{k}$ are the $k$th user's SINRs.
Therefore, the $k$th user's maximum achievable common rate and private rate are written as $R_{\mathrm{c},k}\!(\mathbf{h}_{k}) \! = \! -\log_{2}(\varepsilon_{\mathrm{c},k}^{\mathrm{MMSE}})$ and $R_{k}\!(\mathbf{h}_{k}) \! = \! -\log_{2}(\varepsilon_{k}^{\mathrm{MMSE}})$, respectively.
The common message is transmitted at a common rate defined as $R_{\mathrm{c}} \triangleq \min_{j}\{R_{\mathrm{c},j}\}_{j=1}^{K}$, which ensures that it is decodable by all users.
Ultimately, the objective would be to design $\mathbf{P}$ such that the SR given as $R_{\mathrm{c}} + \sum_{k=1}^{K}R_{k}$ is maximized. It can be seen that in scenarios where the multicast part is not beneficial, allocating zero power to the common precoder will yield $R_{\mathrm{c}} = 0$, and the system reduces to a conventional BC.
However, rates are functions of the actual channel and hence cannot be used to construct an optimization problem at the BS.
Alternatively, we consider the Average Rates (ARs) defined as: $\mathrm{E}_{\mathbf{h}_{k}\mid \widehat{\mathbf{h}}_{k}}\{R_{\mathrm{c},k}\}$ and $\mathrm{E}_{\mathbf{h}_{k}\mid \widehat{\mathbf{h}}_{k}}\{R_{k}\}$. In the following, $\mathrm{E}_{\mathbf{h}_{k}\mid \widehat{\mathbf{h}}_{k}}\{\cdot\}$ will be simply referred to as $\mathrm{E}\{\cdot\}$.
Before we proceed to the ASR problem formulation, we highlight the benefit of incorporating the common symbol from a DoF perspective.
\subsection{DoF Motivated Design}
\label{subsection_DoF_scheme}
The DoF-motivated JMB design in \cite{Hao2013} is briefly revisited in this subsection.
Consider
$\sigma_{n}^{2} = 1 \Rightarrow \mathrm{SNR} = P_{t}$, and an average estimation error power $ \mathrm{E} \big\{ \| \widetilde{\mathbf{h}}_{k} \|^{2} \big\} $
that decays as $O \left( P_{t}^{-\alpha} \right)$, where $\alpha \geq 0$ is an exponent that represents the CSIT quality.
For example, $\alpha =0$ represents a fixed error power w.r.t SNR, e.g. constant number of feedback bits.
On the other hand, $\alpha = \infty$ corresponds to perfect CSIT.
In DoF analysis, it is customary to truncate the exponent such that $\alpha \leq 1$, where $\alpha = 1$ corresponds to perfect CSIT from a DoF perspective \cite{Yang2013}.
Under these assumptions, the precoders of the private symbols are given as
$\mathbf{p}_{k} = \sqrt{P_{t}^{\alpha}/K} \widehat{\mathbf{p}}_{k}^{\mathrm{ZF}}$,
where $\widehat{\mathbf{p}}_{k}^{\mathrm{ZF}}$ is a normalized ZF-BF vector constructed using the channel estimate $\widehat{\mathbf{H}}$, such that $\|\widehat{\mathbf{p}}_{k}^{\mathrm{ZF}}\| = 1$ and $\widehat{\mathbf{h}}_{i}^{H}\widehat{\mathbf{p}}_{k}^{\mathrm{ZF}} = 0$, $\forall i,k \in \mathcal{K}, \ i\neq k$.
The common symbol's precoder is given as $\mathbf{p}_{\mathrm{c}} = \sqrt{P_{t} - P_{t}^{\alpha}}\mathbf{e}_{1}$, where $\mathbf{e}_{1}$ is a standard unity basis vector with $1$ as the first entry and zeros elsewhere.
The $k$th user's received signal is given as
\begin{equation}
\nonumber
  y_{k} =
  \underset{ O  ( P_{t}^{1} )}{\underbrace{\mathbf{h}_{k}^{H} \mathbf{p}_{\mathrm{c}}s_{\mathrm{c}}}}
  + \underset{ O( P_{t}^{\alpha})}{\underbrace{ \mathbf{h}_{k}^{H} \mathbf{p}_{k}s_{k} }}
  +  \sum_{i\neq k}\underset{O( P_{t}^{0} )}{\underbrace{\widetilde{\mathbf{h}}_{k}^{H}\mathbf{p}_{i}s_{i}}}
  + \underset{O ( P_{t}^{0} )}{\underbrace{n_{k}}}
\end{equation}
indicating the order of the average power of each term as $P_{t} \rightarrow \infty$.
The third Right Hand Side (RHS) term corresponds to the residual interference from unintended private symbols, resulting from the employment of an imperfect channel estimate to construct the ZF-BF vectors.
Since the error scales as $O \left( P_{t}^{-\alpha} \right)$ whilst the power allocated to the private precoders scales as $O \left( P_{t}^{\alpha} \right)$, the residual interference is drowned by noise and can be neglected.
By decoding the common symbol while treating the rest of the terms as noise, a DoF of $1-\alpha$ is achieved.
Moreover, the private symbol achieves a DoF of $\alpha$ after cancelling the common symbol.
The same applies to the other users, and a sum DoF of $1+(K-1)\alpha$ is achieved.
On the other hand, excluding the common symbol and splitting $P_{t}$ between the private symbols, the receive power of the intended private symbol is enhanced to $O( P_{t}^{1})$.
However, residual interference is also increased to $O( P_{t}^{1-\alpha})$, and the sum DoF obtained by the private symbols remains as $K\alpha$.
Therefore, JMB is strictly superior to ZF-BF and SU transmission (e.g. TDMA which achieves a DoF of 1) for $ 0 < \alpha < 1$.  The reader is referred to \cite{Yang2013,Hao2013} for more on  DoF analysis.
It is clear that the DoF-motivated scheme adapts to the CSIT accuracy by changing the power allocation.
However, the fact that the precoders are not optimized leaves considerable potential for improvement, particularly in the finite SNR regime.
\subsection{Problem Formulation}
In order to formulate a deterministic ASR problem, the stochastic ARs are approximated  by corresponding Sample Average Functions (SAFs).
Each SAF is obtained  by taking the ensemble average over a sample of $M$ independent identically distributed (i.i.d) realizations drawn from the distribution $f_{\mathbf{H}|\widehat{\mathbf{H}}}$ in a Monte-Carlo fashion.
The sample is defined as
$\mathbf{H}_{\mathcal{M}} \triangleq \left\{ \mathbf{H}^{(m)} \mid m \in \mathcal{M} \right\}$,
where $\mathbf{H}^{(m)} \triangleq [\mathbf{h}_{1}^{(m)} ,\ldots,\mathbf{h}_{K}^{(m)}]$ is the $m$th realization, and $\mathcal{M} \triangleq \left\{1,\ldots,M\right\}$.
The SAFs are given as: $\bar{R}_{\mathrm{c},k}^{(M)} = \frac{1}{M} \sum_{m=1}^{M} R_{\mathrm{c},k}^{(m)} $ and $\bar{R}_{k}^{(M)} =\frac{1}{M} \sum_{m=1}^{M} R_{k}^{(m)}$, where
$R_{\mathrm{c},k}^{(m)} \triangleq R_{\mathrm{c},k}\big(\mathbf{h}_{k}^{(m)}\big) $ and
$R_{k}^{(m)} \triangleq R_{k}\big(\mathbf{h}_{k}^{(m)}\big)$
are the rates associated with the realization $\mathbf{h}_{k}^{(m)}$.
In the following, the superscript $(m)$ is used to indicate the association of variables with the
$m$th Monte-Carlo realization.
It should be noted that $\mathbf{P}$ is fixed over the $M$ realizations of the rates, which follows from the definition of the ARs.
This also reflects the fact that $\mathbf{P}$ is optimized at the BS using partial CSI knowledge.
\newtheorem{Assumption_SNR}[Assumption_Counter]{Assumption}
\begin{Assumption_SNR}
\label{Assumption_SNR}
\textnormal{
In the following, we assume that $\frac{\sigma_{n}^{2}}{\|\mathbf{h}_{k}\|^{2}P_{t}} > 0$ with probability 1, $\forall k \in \mathcal{K}$.
}
\end{Assumption_SNR}
Alternatively, we can say that $\mathrm{SNR} = {P_{t}}/{\sigma_{n}^{2}}$ can only grow finitely large, and channel gains are finite.
Assumption \ref{Assumption_SNR} yields $\varepsilon_{\mathrm{c},k}^{\mathrm{MMSE}},\varepsilon_{k}^{\mathrm{MMSE}} > 0$ with probability $1$, as the presence of a nonzero noise variance dictates that $E_{\mathrm{c},k},E_{k}>0$.
This also implies that rates are finite, and by the strong law of large numbers we can write
\vspace{-2.0mm}
\begin{subequations}
\label{Eq_R_LLN}
\begin{align}
\label{Eq_R_LLN_c}
  \bar{R}_{\mathrm{c},k} \triangleq & \  \lim_{M\rightarrow \infty} \bar{R}_{\mathrm{c},k}^{(M)} = \mathrm{E} \{ R_{\mathrm{c},k} \},
  \text{almost surely} \\
  \label{Eq_R_LLN_p}
  \bar{R}_{k} \triangleq & \  \lim_{M\rightarrow \infty} \bar{R}_{k}^{(M)} = \mathrm{E} \{ R_{k} \},
  \text{almost surely}
\end{align}
\end{subequations}
where $\bar{R}_{\mathrm{c},k}$ and $\bar{R}_{k}$ are the approximated ARs for a sufficiently large $M$, which will be simply referred to as the ARs. The common AR is defined as $\bar{R}_{\mathrm{c}} \triangleq \min_{j}\{\bar{R}_{\mathrm{c},j}\}_{j=1}^{K}$.
The objective is to design $\mathbf{P}$ that maximizes the ASR defined as $\bar{R}_{\mathrm{c}}  + \sum_{k=1}^{K} \bar{R}_{k}$, subject to a power constraint $P_{t}$.
This problem is formulated as
\vspace{-2.0mm}
\begin{subequations}
 \label{Eq_Opt_ASR_M}
\begin{align}
\label{Eq_Opt_ASR_M_a0}
\overline{\bm{\mathcal{R}}}:  \underset{\bar{R}_{\mathrm{c}}, \mathbf{P} }{\max} \ &
 \bar{R}_{\mathrm{c}} + \sum_{k=1}^{K} \bar{R}_{k}   \\
 \label{Eq_Opt_ASR_M_a}
 \text{s.t.} \quad \; &  \bar{R}_{\mathrm{c},k} \geq \bar{R}_{\mathrm{c}}, \; \forall k\in\mathcal{K} \\
 \label{Eq_Opt_ASR_M_b}
  &  \mathrm{tr}\big(\mathbf{P}\mathbf{P}^{H}\big) \leq P_{t}
\end{align}
\end{subequations}
where the constraints in \eqref{Eq_Opt_ASR_M_a} are introduced to eliminate the potential non-smoothness arising from the pointwise minimization in $\bar{R}_{\mathrm{c}}$.
Problem $\overline{\bm{\mathcal{R}}}$ is a non-convex optimization problem that appears to be very challenging to solve.
\section{AWSMSE Optimization}
\label{Section_ASR_AWSMSE_Optimization}
In this section, the ASR problem is transformed into an equivalent problem that can be solved using AO.
We start by introducing the main components used to construct the equivalent problem, i.e. the augmented WMSEs \cite{Shi2011}:
\vspace{-1.0mm}
\begin{subequations}
\label{Eq_A_WMSEs}
\begin{align}
\label{Eq_A_WMSEs_c}
\xi_{\mathrm{c},k}\big( \! \mathbf{h}_{k},g_{\mathrm{c},k},u_{\mathrm{c},k} \! \big) & =  u_{\mathrm{c},k}(\mathbf{h}_{k})\varepsilon_{\mathrm{c},k}(\mathbf{h}_{k}) \! - \! \log_{2} \! \big( \! u_{\mathrm{c},k}(\mathbf{h}_{k}) \! \big) \\
\label{Eq_A_WMSEs_p}
\xi_{k}\big(\mathbf{h}_{k},g_{k},u_{k}\big) & =  u_{k}(\mathbf{h}_{k})\varepsilon_{k}(\mathbf{h}_{k}) - \log_{2}\big(u_{k}(\mathbf{h}_{k})\big)
\end{align}
\end{subequations}
where $u_{\mathrm{c},k}(\mathbf{h}_{k})\geq 0$ and $ u_{k}(\mathbf{h}_{k}) \geq 0$ are weights associated with the $k$th user's MSEs, and the dependencies in \eqref{Eq_A_WMSEs} are highlighted for their significance in the following analysis.
The dependencies of the weights on the actual channel is crucial for the establishment of the following WMSE-Rate relationship:
\vspace{-1.0mm}
\begin{equation}
\label{Eq_min_WMSE}
 \underset{u_{\mathrm{c},k}, g_{\mathrm{c},k}}{\min} \xi_{\mathrm{c},k} = 1-R_{\mathrm{c},k}
 \quad \text{and} \quad
 \underset{u_{k}, g_{k}}{\min} \ \xi_{k}= 1-R_{k}.
\end{equation}
This can be shown as follows: from
$\frac{\partial\xi_{\mathrm{c},k}}{\partial g_{\mathrm{c},k}} = 0$
and
$\frac{\partial\xi_{k}}{\partial g_{k}} = 0$, the optimum equalizers are obtained as
$g_{\mathrm{c},k}^{\ast}  = g_{\mathrm{c},k}^{\mathrm{MMSE}} $ and $g_{k}^{\ast} = g_{k}^{\mathrm{MMSE}}$.
Substituting this back into \eqref{Eq_A_WMSEs}, we obtain the augmented WMMSEs written as
\vspace{-1.0mm}
\begin{subequations}
\label{Eq_A_WMMSEs}
\begin{align}
\xi_{\mathrm{c},k}^{\mathrm{MMSE}}( \mathbf{h}_{k},u_{\mathrm{c},k}  \big) & =  u_{\mathrm{c},k}\varepsilon_{\mathrm{c},k}^{\mathrm{MMSE}} - \log_{2}(u_{\mathrm{c},k}) \\
\xi_{k}^{\mathrm{MMSE}}( \mathbf{h}_{k},u_{k} \big) & =  u_{k}\varepsilon_{k}^{\mathrm{MMSE}} - \log_{2}(u_{k}).
\end{align}
\end{subequations}
Furthermore, from
$\frac{\partial\xi_{\mathrm{c},k}^{\mathrm{MMSE}}}{\partial u_{\mathrm{c},k}} = 0$
and
$\frac{\partial\xi_{k}^{\mathrm{MMSE}}}{\partial u_{k}} = 0$, we obtain the optimum MMSE weights:
$u_{\mathrm{c},k}^{\ast} = u_{\mathrm{c},k}^{\mathrm{MMSE}} \triangleq \big( \varepsilon_{\mathrm{c},k}^{\mathrm{MMSE}} \big)^{-1}$
and
$u_{k}^{\ast} = u_{k}^{\mathrm{MMSE}} \triangleq \big( \varepsilon_{k}^{\mathrm{MMSE}} \big)^{-1}$.
Substituting this back into \eqref{Eq_A_WMMSEs} yields the relationship in \eqref{Eq_min_WMSE}.
It is evident from \eqref{Eq_MMSE} that the MMSE weights are dependent on the channel.

The equivalent problem is formulated using the augmented AWMSEs defined as:
$\mathrm{E}\{\xi_{\mathrm{c},k}\} $ and $\mathrm{E}\{\xi_{k}\}$.
Before we proceed, the augmented AWMSEs are approximated as:
%
\vspace{-2.0mm}
%
\begin{align}
\nonumber
\bar{\xi}_{\mathrm{c},k}^{(M)} & =  \frac{1}{M} \sum_{m=1}^{M}\xi_{\mathrm{c},k}^{(m)}
\quad \text{and} \quad
\bar{\xi}_{k}^{(M)} = \frac{1}{M} \sum_{m=1}^{M}\xi_{k}^{(m)}, \ \text{where} \\
\nonumber
\xi_{\mathrm{c},k}^{(m)} \! & \triangleq  \xi_{\mathrm{c},k}\big(  \mathbf{h}_{k}^{(m)} \! \!,g_{\mathrm{c},k}^{(m)} \! \!,u_{\mathrm{c},k}^{(m)}  \big)
\ \text{and} \
\xi_{k}^{(m)} \! \triangleq  \xi_{k}\big(  \mathbf{h}_{k}^{(m)} \! \!,g_{k}^{(m)} \! \!,u_{k}^{(m)}  \big)
\end{align}
correspond to the $m$th realization of the augmented WMSEs,
which depend on the $m$th realization of the equalizers:
$g_{\mathrm{c},k}^{(m)} \!  \triangleq \!  g_{\mathrm{c},k}\big( \!  \mathbf{h}_{k}^{(m)} \!  \big)  $
and
$g_{k}^{(m)} \!  \triangleq \!  g_{k}\big( \!  \mathbf{h}_{k}^{(m)} \!  \big)  $,
and the weights:
$u_{\mathrm{c},k}^{(m)} \!  \triangleq \!  u_{\mathrm{c},k}\big( \!  \mathbf{h}_{k}^{(m)} \!  \big)  $
and
$u_{k}^{(m)} \!  \triangleq \!  u_{k}\big(\! \mathbf{h}_{k}^{(m)} \!  \big)  $.
For compactness, we define the set of equalizers associated with the $M$ realizations and the $K$ users as:
$\mathbf{G} \!  \triangleq \!   \big\{ \!  \mathbf{g}_{\mathrm{c},k},\mathbf{g}_{k} \!  \mid \!  k \!  \in \!  \mathcal{K} \!  \big\}$,
where
$\mathbf{g}_{\mathrm{c},k} \!  \triangleq \!  \big\{ \!  g_{\mathrm{c},k}^{(m)} \!  \mid \!  m \!  \in \!  \mathcal{M} \!  \big\}$
and
$\mathbf{g}_{k} \!  \triangleq \!  \big\{ \!  g_{k}^{(m)} \!  \mid \!  m  \! \in \!  \mathcal{M} \!  \big\}$.
In a similar manner, we define:
$\mathbf{U} \!  \triangleq \!   \big\{ \!  \mathbf{u}_{\mathrm{c},k},\mathbf{u}_{k} \!  \mid \!  k \!  \in \!  \mathcal{K} \!  \big\}$,
where
$\mathbf{u}_{\mathrm{c},k} \!  \triangleq \!  \big\{ \!  u_{\mathrm{c},k}^{(m)} \!  \mid \!  m \!  \in \!  \mathcal{M} \!  \big\}$
and
$\mathbf{u}_{k} \!  \triangleq \!  \big\{ \!  u_{k}^{(m)} \!  \mid \!  m \!  \in \!  \mathcal{M} \!  \big\}$.
The approximated augmented AWMSEs for a sufficiently large $M$ are defined as
$\bar{\xi}_{\mathrm{c},k} \!  \triangleq \!  \lim_{M\rightarrow \infty} \!  \bar{\xi}_{\mathrm{c},k}^{(M)}$
and
$\bar{\xi}_{k} \!  \triangleq \!  \lim_{M\rightarrow \infty} \!  \bar{\xi}_{k}^{(M)}$,
which will be simply referred to as the AWMSEs.
The same approach used to prove \eqref{Eq_min_WMSE} can be employed to show that
\vspace{-1.0mm}
\begin{equation}
\label{Eq_min_AWMSE}
 \underset{\mathbf{u}_{\mathrm{c},k}, \mathbf{g}_{\mathrm{c},k}}{\min} \bar{\xi}_{\mathrm{c},k} = 1-\bar{R}_{\mathrm{c},k}
 \quad \text{and} \quad
 \underset{\mathbf{u}_{k}, \mathbf{g}_{k}}{\min} \ \bar{\xi}_{k}= 1-\bar{R}_{k}
\end{equation}
where optimality conditions are checked separately for each realization.
The sets of optimum MMSE equalizers associated with \eqref{Eq_min_AWMSE} are defined as
$\mathbf{g}^{\mathrm{MMSE}}_{\mathrm{c},k} \! \triangleq \! \big\{ \! g_{\mathrm{c},k}^{\mathrm{MMSE}(m)}  \! \mid \! m \!  \in \!  \mathcal{M} \! \big\}$
and
$\mathbf{g}^{\mathrm{MMSE}}_{k} \! \triangleq \! \big\{ \! g_{k}^{\mathrm{MMSE}(m)} \! \mid \! m \!  \in \!  \mathcal{M} \!  \big\}$.
In the same manner, the sets of optimum MMSE weights are defined as
$\mathbf{u}^{\mathrm{MMSE}(m)}_{\mathrm{c},k} \! \triangleq \! \big\{ \! u_{\mathrm{c},k}^{\mathrm{MMSE}(m)} \! \mid \! m \!  \in \!  \mathcal{M} \! \big\}$
and
$\mathbf{u}^{\mathrm{MMSE}(m)}_{k}  \! \triangleq \! \big\{ \! u_{k}^{\mathrm{MMSE}(m)} \! \mid \! m \!  \in \!  \mathcal{M} \! \big\}$.
For the $K$ users, the MMSE solution is composed as
$\mathbf{G}^{\mathrm{MMSE}} \! \triangleq \!  \big\{ \! \mathbf{g}_{\mathrm{c},k}^{\mathrm{MMSE}}, \mathbf{g}_{k}^{\mathrm{MMSE}} \! \mid \! k \!  \in \!  \mathcal{K} \! \big\}$
and
$\mathbf{U}^{\mathrm{MMSE}} \! \triangleq  \! \big\{ \! \mathbf{u}_{\mathrm{c},k}^{\mathrm{MMSE}}, \mathbf{u}_{k}^{\mathrm{MMSE}} \! \mid \! k  \! \in \!  \mathcal{K} \! \big\}$.
\subsection{Augmented AWSMSE Minimization}
Motivated by the relationship in \eqref{Eq_min_AWMSE}, the augmented AWSMSE minimization problem is formulated as
\vspace{-2.0mm}
\begin{subequations}
\label{Eq_Opt_AWSMSE_M}
\begin{align}
\label{Eq_Opt_AWSMSE_M_a_0}
\overline{\bm{\mathcal{A}}}:  \underset{\bar{\xi}_{\mathrm{c}}, \mathbf{P},\mathbf{U},\mathbf{G} }{\min} \ &
\bar{\xi}_{\mathrm{c}} + \sum_{k=1}^{K} \bar{\xi}_{k}   \\
\label{Eq_Opt_AWSMSE_M_a}
\text{s.t.} \quad \; &  \bar{\xi}_{\mathrm{c},k} \leq \bar{\xi}_{\mathrm{c}}, \; \forall k\in\mathcal{K} \\
\label{Eq_Opt_AWSMSE_M_b}
 &   \mathrm{tr}\big(\mathbf{P}\mathbf{P}^{H}\big) \leq P_{t}
\end{align}
\end{subequations}
where $\bar{\xi}_{\mathrm{c}}$ is an auxiliary variable.
The MMSE solution of the problems in \eqref{Eq_min_AWMSE} is not only optimum for problem $\overline{\bm{\mathcal{A}}}$,
but it is also at the heart of a relationship that connects the stationary points of problem $\overline{\bm{\mathcal{A}}}$ to the stationary points of problem $\overline{\bm{\mathcal{R}}}$.
In the following, the notations $\mathbf{G}^{\mathrm{MMSE}}(\mathbf{P})$ and $\mathbf{U}^{\mathrm{MMSE}}(\mathbf{P})$ are used
to emphasize the dependencies on  particular precoders.
\newtheorem{Proposition_ASR_M_AWSMSE_M}[Proposition_Counter]{Proposition}
\begin{Proposition_ASR_M_AWSMSE_M}\label{Proposition_ASR_M_AWSMSE_M}
\textnormal{
For any stationary point of $\overline{\bm{\mathcal{A}}}$ given as
$\left(\bar{\xi}_{\mathrm{c}}^{\ast},\mathbf{P}^{\ast},\mathbf{U}^{\ast},\mathbf{G}^{\ast}\right)$ that achieves an objective function of $\bar{\xi}^{\ast}$, there exists:
1) a corresponding MMSE stationary point given as
$\left(\bar{\xi}_{\mathrm{c}}^{\ast},\mathbf{P}^{\ast},\mathbf{U}^{\mathrm{MMSE}}(\mathbf{P}^{\ast}),\mathbf{G}^{\mathrm{MMSE}}(\mathbf{P}^{\ast})\right)$
that achieves the same objective function,
2) a stationary point of $\overline{\bm{\mathcal{R}}}$ given as
$\left(1-\bar{\xi}_{\mathrm{c}}^{\ast},\mathbf{P}^{\ast}\right)$ that achieves an ASR of $K+1 - \bar{\xi}^{\ast}$.
Finally, if $\left(\bar{\xi}_{\mathrm{c}}^{\ast},\mathbf{P}^{\ast},\mathbf{U}^{\ast},\mathbf{G}^{\ast}\right)$ is a global optimal point of  $\overline{\bm{\mathcal{A}}}$, then $\left(1-\bar{\xi}_{\mathrm{c}}^{\ast},\mathbf{P}^{\ast}\right)$ must be a global optimal point of $\overline{\bm{\mathcal{R}}}$.
}
\end{Proposition_ASR_M_AWSMSE_M}
This can be shown by employing the ideas used to prove \cite[Proposition 1]{Razaviyayn2013b}. A sketch of the proof is given as follows.
\begin{proof}[Proof of Proposition \ref{Proposition_ASR_M_AWSMSE_M}]
From the KKT conditions of $\overline{\bm{\mathcal{A}}}$, it can be seen that
$\big( \mathbf{g}_{k}^{\mathrm{MMSE}} \!,\mathbf{u}_{k}^{\mathrm{MMSE}} \big)$ is optimal and unique $\forall k \in \mathcal{K}$.
Moreover,
$\big( \mathbf{g}_{\mathrm{c},k}^{\mathrm{MMSE}}\!,\mathbf{u}_{\mathrm{c},k}^{\mathrm{MMSE}}  \big)$ is optimal and unique $\forall k \in \mathcal{K}_{\mathrm{A}}$, where $\mathcal{K}_{\mathrm{A}} \subseteq \mathcal{K}$ is the set of active constraints in \eqref{Eq_Opt_AWSMSE_M_a}.
For inactive constraints, the MMSE solution is not unique but it satisfies the optimality conditions. This proves the first part.
The second part is proved by examining the KKT conditions of problem $\overline{\bm{\mathcal{R}}}$ and employing the relationship in \eqref{Eq_min_AWMSE}. The final part is proved by contradiction.
For the complete proof, readers are referred to the extended version of this paper \cite{Joudeh2014a}.
\end{proof}
%
\section{Alternating Optimization Algorithm}
\label{Section_AO}
Although problem $\overline{\bm{\mathcal{A}}}$ is non-convex in the joint set of optimization variables, it is convex  in each of the blocks $\mathbf{P}$, $\mathbf{U}$ and $\mathbf{G}$, assuming that the other two are fixed.
This block-wise convexity is exploited using an AO algorithm that switches between optimizing blocks.
Each iteration of the proposed algorithm consists of two steps: 1) updating $\mathbf{G}$ and $\mathbf{U}$ for a given $\mathbf{P}$,
2) updating $\mathbf{P}$ for given $\mathbf{G}$ and $\mathbf{U}$.
\subsection{Updating the Equalizers and Weights}
\label{subsection_obj_fn_update}
In $n$th iteration of the AO algorithm, the equalizers and weights are updated such that
$\mathbf{G} = \mathbf{G}^{\mathrm{MMSE}}\big(\ddot{\mathbf{P}}\big)$ and $\mathbf{U} = \mathbf{U}^{\mathrm{MMSE}}\big(\ddot{\mathbf{P}}\big)$ respectively,
where $\ddot{\mathbf{P}}$ is the precoding matrix obtained in the $(n-1)$th iteration.
To facilitate the problem formulation in the following step, the AWMSEs are written in terms of the updated blocks $\mathbf{G}$ and $\mathbf{U}$, and the block $\mathbf{P}$ which is yet to be update. For this purpose, we introduce the AWMMSE-components listed as:
$\bar{\mathbf{\Psi}}_{\mathrm{c},k}$, $\bar{\mathbf{\Psi}}_{k}$, $\bar{t}_{\mathrm{c},k}$, $\bar{t}_{k}$,
$\bar{\mathbf{f}}_{\mathrm{c},k}$, $\bar{\mathbf{f}}_{k}$, $\bar{u}_{\mathrm{c},k}$, $\bar{u}_{k}$,
$\bar{\upsilon}_{\mathrm{c},k}$ and $\bar{\upsilon}_{k}$, which are obtained using the updated $\mathbf{G}$ and $\mathbf{U}$.
In particular, the components $\bar{u}_{\mathrm{c},k}$ and $\bar{u}_{k}$ are calculated by taking the ensemble averages over the $M$ realizations of
$u_{\mathrm{c},k}^{(m)}$ and $u_{k}^{(m)}$.
The rest of the components are calculated in a similar manner by averaging over their corresponding realizations given as:
%
\vspace{-2.0mm}
\begin{align}
\nonumber
t_{\mathrm{c},k}^{(m)}& = u_{\mathrm{c},k}^{(m)}\left| g_{\mathrm{c},k}^{(m)}\right|^{2}
&\text{and} &&
t_{k}^{(m)} & =  u_{k}^{(m)}\left| g_{k}^{(m)}\right|^{2}\\
\nonumber
\mathbf{\Psi}_{\mathrm{c},k}^{(m)} & =  t_{\mathrm{c},k}^{(m)} \mathbf{h}_{k}^{(m)}{\mathbf{h}_{k}^{(m)}}^{H}
&\text{and} &&
\mathbf{\Psi}_{k}^{(m)} & =   t_{k}^{(m)} \mathbf{h}_{k}^{(m)}{\mathbf{h}_{k}^{(m)}}^{H}\\
\nonumber
\mathbf{f}_{\mathrm{c},k}^{(m)} & = u_{\mathrm{c},k}^{(m)} \mathbf{h}_{k}^{(m)}{g_{\mathrm{c},k}^{(m)}}^{H}
&\text{and} &&
\mathbf{f}_{k}^{(m)} & =  u_{k}^{(m)} \mathbf{h}_{k}^{(m)}{g_{k}^{(m)}}^{H} \\
\nonumber
\upsilon_{\mathrm{c},k}^{(m)} & =  \log_{2}\left(u_{\mathrm{c},k}^{(m)}\right)
&\text{and} &&
\upsilon_{k}^{(m)} & =  \log_{2}\left(u_{k}^{(m)}\right).
\end{align}
The AWMSEs are written in terms of the updated $(\mathbf{G},\mathbf{U})$, and $\mathbf{P}$ which is yet to be updated, as
\vspace{-2.0mm}
\begin{subequations}
\label{Eq_AWMSE}
\begin{align}
 \nonumber
  \bar{\xi}_{\mathrm{c},k}&  =  \mathbf{p}_{\mathrm{c}}^{H} \bar{\mathbf{\Psi}}_{\mathrm{c},k} \mathbf{p}_{\mathrm{c}}
    + \sum_{i=1}^{K}  \mathbf{p}_{i}^{H} \bar{\mathbf{\Psi}}_{\mathrm{c},k} \mathbf{p}_{i}
    + \sigma_{n}^{2}\bar{t}_{\mathrm{c},k}
    - 2\Re \big\{  \bar{\mathbf{f}}_{\mathrm{c},k}^{H} \mathbf{p}_{\mathrm{c}}  \big\}
    \\
    & \ +\bar{u}_{\mathrm{c},k} - \bar{\upsilon}_{\mathrm{c},k}
    \\
   \bar{\xi}_{k} &  =   \sum_{i=1}^{K}  \mathbf{p}_{i}^{H} \bar{\mathbf{\Psi}}_{k} \mathbf{p}_{i}
      +  \sigma_{n}^{2}\bar{t}_{k}
      -  2\Re \big\{ \bar{\mathbf{f}}_{k}^{H} \mathbf{p}_{k}\big\}
      +  \bar{u}_{k}  \!  -  \!  \bar{\upsilon}_{k}.
\end{align}
\end{subequations}
\subsection{Updating the Precoders}
\label{Subsection_Opt Pre_Vect}
Following the previous step, the problem of updating $\mathbf{P}$ is denoted by $\overline{\bm{\mathcal{A}}}_{\mathbf{P}}$, which
is formulated by substituting \eqref{Eq_AWMSE} into \eqref{Eq_Opt_AWSMSE_M} and eliminating $(\mathbf{G},\mathbf{U})$ from the set of optimization variables. This is given as
\vspace{-2.0mm}
\begin{subequations}
 \label{Eq_WAMSE_QCQP}
\begin{align}
\label{Eq_WAMSE_QCQP_a_0}
\overline{\bm{\mathcal{A}}}_{\mathbf{P}}: &
\underset{\bar{\xi}_{\mathrm{c}} ,  \mathbf{P}}{\min} \   \bar{\xi}_{\mathrm{c}} + \sum_{k=1}^{K}  \bigg(
      \sum_{i=1}^{K}
      \mathbf{p}_{i}^{H}\bar{\mathbf{\Psi}}_{k}\mathbf{p}_{i}
    - 2\Re\big\{\bar{\mathbf{f}}_{k}^{H} \mathbf{p}_{k}\big\}
      \bigg) \\
 \nonumber
      \text{s.t.} \ &
      \mathbf{p}_{\mathrm{c}}^{H}\bar{\mathbf{\Psi}}_{\mathrm{c},k}\mathbf{p}_{\mathrm{c}}
    \! + \! \sum_{i=1}^{K}
     \mathbf{p}_{i}^{H}\bar{\mathbf{\Psi}}_{\mathrm{c},k}\mathbf{p}_{i}
    \! + \! \sigma_{n}^{2}\bar{t}_{\mathrm{c},k}
    \!- \! 2 \Re \big\{ \bar{\mathbf{f}}_{\mathrm{c},k}^{H} \mathbf{p}_{\mathrm{c}} \big\}
    \\
    \label{Eq_WAMSE_QCQP_a}
    & \quad
     +  \bar{u}_{\mathrm{c},k}
     -  \bar{\upsilon}_{\mathrm{c},k}
    \leq \bar{\xi}_{\mathrm{c}}, \ \forall k\in \mathcal{K} \\
 \label{Eq_WAMSE_QCQP_b}
                        &  \quad
                         \mathrm{tr}\big(\ \mathbf{P} \mathbf{P}^{H} \big) \leq P_{t}
\end{align}
\end{subequations}
where the constant term $\sum_{k=1}^{K}(\sigma_{n}^{2}\bar{t}_{k} + \bar{u}_{k}-\bar{\upsilon}_{k})$ has been omitted from \eqref{Eq_WAMSE_QCQP_a_0}.
Problem \eqref{Eq_WAMSE_QCQP} is a convex Quadratically Constrained Quadratic Program (QCQP) which can be solved using interior-point methods \cite{Grant2008}.
\subsection{Alternating Optimization Algorithm}
\label{Subsection_AO}
The AO algorithm is constructed by repeating the steps described in the two previous subsections until convergence.
This is summarized in Algorithm \ref{Algthm_AO} where $\epsilon_{R}$ determines the accuracy of the solution and $n_{\max}$ is the maximum number of iterations. Step \ref{Algthm_AO_step_initialize} is discussed in Section \ref{Section_Numerical_Results}.
%
\begin{algorithm}
\caption{Alternating Optimization}
\label{Algthm_AO}
\begin{algorithmic}[1]
\State \textbf{Initialize}: $n\gets 0$, $\bar{R}^{(n)} \gets 0$, $\mathbf{P}$
\label{Algthm_AO_step_initialize}
\Repeat
    \State $n\gets n+1$, $\ddot{\mathbf{P}}\gets \mathbf{P}$
    \State $\mathbf{G}\gets \mathbf{G}^{\mathrm{MMSE}}\big(\ddot{\mathbf{P}}\big)$, $\mathbf{U}\gets \mathbf{U}^{\mathrm{MMSE}}\big(\ddot{\mathbf{P}}\big)$
    \State update
    $\big\{\bar{\mathbf{\Psi}}_{\mathrm{c},k},\bar{\mathbf{\Psi}}_{k},\bar{\mathbf{F}}_{\mathrm{c},k},\bar{\mathbf{F}}_{k}, \bar{\mathbf{t}}_{\mathrm{c},k},\bar{\mathbf{u}}_{\mathrm{c},k},\bm{\bar{\upsilon}}_{\mathrm{c},k},\bm{\bar{\upsilon}}_{k}\big\}_{k=1}^{K}$
    \label{Algthm_AO_step_expectations}
     \State  $\mathbf{P} \gets \arg \overline{\bm{\mathcal{A}}}_{\mathbf{P}}$
    \label{Algthm_AO_step_SDP}
    \State $\bar{R}^{(n)} \gets \underset{j}{\min}\{ \bar{\upsilon}_{\mathrm{c},j} \}_{j=1}^{K} + \sum_{k=1}^{K}\bar{\upsilon}_{k}$
\Until{$\left|\bar{R}^{(n)} - \bar{R}^{(n-1)} \right| < \epsilon_{R}$ \text{or} $n=n_{\max}$ }
\end{algorithmic}
\end{algorithm}
\newtheorem{Proposition_WAMSE_Conv}[Proposition_Counter]{Proposition}
\begin{Proposition_WAMSE_Conv}\label{Proposition_WAMSE_Conv}
\textnormal{
Algorithm \ref{Algthm_AO} converges to a stationary point of problem $\overline{\bm{\mathcal{A}}}$ denoted by
$\big(\bar{\xi}_{\mathrm{c}}^{\ast},\mathbf{P}^{\ast},\mathbf{U}^{\ast},\mathbf{G}^{\ast} \big)$,
where the corresponding $\mathbf{P}^{\ast}$ is a stationary solution of problem $\overline{\bm{\mathcal{R}}}$.
}
\end{Proposition_WAMSE_Conv}
This can be proved by employing the ideas in \cite[Theorem 2]{Razaviyayn2013b}. A sketch of the proof is given as follows.
\begin{proof}[Proof of Proposition \ref{Proposition_WAMSE_Conv}]
The iterations of Algorithm \ref{Algthm_AO} monotonically decrease the cost function of $\overline{\bm{\mathcal{A}}}$.
Moreover, the feasibility set in constraint \eqref{Eq_Opt_AWSMSE_M_b} is compact, and the mappings
$\mathbf{G}^{\mathrm{MMSE}}(\mathbf{P})$ and $\mathbf{U}^{\mathrm{MMSE}}(\mathbf{P})$ are continuous.
Therefore, the iterations converge to a limit point denoted by
$(\mathbf{P}^{\ast},\mathbf{U}^{\ast},\mathbf{G}^{\ast})$,
where $\mathbf{P}^{\ast} = \arg \overline{\bm{\mathcal{A}}}_{\mathbf{P}}(\mathbf{U}^{\ast},\mathbf{G}^{\ast})$,
$\mathbf{U}^{\ast} = \mathbf{U}^{\mathrm{MMSE}}(\mathbf{P}^{\ast})$ and
$\mathbf{G}^{\ast} = \mathbf{G}^{\mathrm{MMSE}}(\mathbf{P}^{\ast})$.
Furthermore, each of the blocks $\mathbf{P}^{\ast}$, $\mathbf{U}^{\ast}$ and $\mathbf{G}^{\ast}$ satisfies the KKT conditions of its corresponding optimization problem, formulated by fixing the other two blocks in $\overline{\bm{\mathcal{A}}}$.
This can be used to show that the point $(\mathbf{P}^{\ast},\mathbf{U}^{\ast},\mathbf{G}^{\ast})$ satisfies the KKT conditions of problem $\overline{\bm{\mathcal{A}}}$.
Combining this with Proposition \ref{Proposition_ASR_M_AWSMSE_M} completes the proof.
\end{proof}
\section{Numerical Results}
\label{Section_Numerical_Results}
\begin{figure}[t!]\vspace{-3mm}
\centering
\includegraphics[width = 0.45\textwidth]{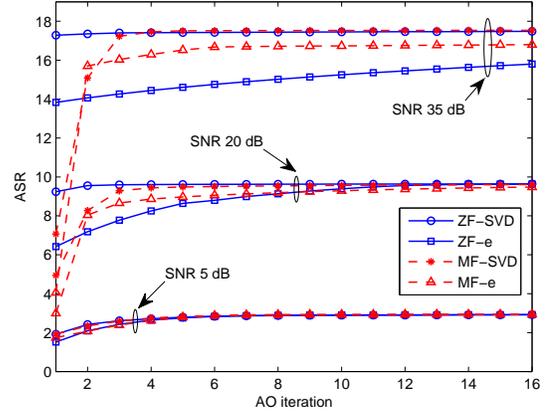}\\
\caption{ASR convergence of Algorithm \ref{Algthm_AO} using 4 different initialization for 1 randomly generated $\mathbf{H}$, $\alpha=0.6$, $\mathrm{SNR} = 5$, $20$ and $35$ dB, and $K=2$.}
\label{Fig_Convergence}
\vspace{-5mm}
\end{figure}
We consider a  MU-MISO system with $N_{t} = K =2$.
Uncorrelated channel fading is assumed, where the entries of $\mathbf{H}$ have a complex Gaussian distribution $\mathcal{C}\mathcal{N}\left(0,1\right)$.
Moreover, the noise variance is fixed as $\sigma_{n}^{2} = 1$, from which the long-term SNR is given as $\mathrm{SNR} = P_{t}$.
Gaussian CSIT error is assumed where the entries of $\widetilde{\mathbf{H}}$ are generated according to the distribution $\mathcal{C}\mathcal{N}\left(0,\sigma_{e}^{2}\right)$.
The error variance is given as $\sigma_{e}^{2} = P_{t}^{-\alpha}$, corresponding to scenarios where the CSIT error decays as SNR increases \cite{Yang2013,Hao2013}.
The value of $\alpha$ is  varied throughout the simulations to represent different CSIT accuracies.
For each realization $\mathbf{H}$, a channel estimation error $\widetilde{\mathbf{H}}$ is drawn from $\mathcal{C}\mathcal{N}\left(0,\sigma_{e}^{2}\right)$, from which the channel estimate is calculated as $\widehat{\mathbf{H}} = \mathbf{H} - \widetilde{\mathbf{H}}$.
A channel realization $\mathbf{H}$ should not be confused with a Monte-Carlo realization $\mathbf{H}^{(m)}$.
While the former is unique for a given transmission and not known to the BS, the latter is part of a sample $\mathbf{H}_{\mathcal{M}}$ generated at the BS in order to formulate the optimization problem.
The size of the sample is set to $M=1000$ throughout the simulations. For a given channel estimate, the $m$th Monte-Carlo realization is obtained as $\mathbf{H}^{(m)}  = \widehat{\mathbf{H}} +  \widetilde{\mathbf{H}}^{(m)}$, where $\widetilde{\mathbf{H}}^{(m)}$ is drawn from the error distribution.

First, we examine the convergence of Algorithm \ref{Algthm_AO} using four different $\mathbf{P}$ initializations.
For the first initialization (ZF-SVD), the precoders of the private messages are initialized as ZF-BF constructed using $\widehat{\mathbf{H}}$, while the common precoder is constructed by taking the dominant left singular vector of $\widehat{\mathbf{H}}$.
The second initialization (ZF-e) uses the standard basis vector $\mathbf{e}_{1}$ to initialize the common precoder.
The third (MF-SVD) and fourth (MF-e) initializations apply Matched Beamformers (M-BF) instead of the ZF-BF.
All initializations use the DoF-motivated power allocation from Section \ref{subsection_DoF_scheme}.
The ASR convergence of the proposed algorithm for $\alpha = 0.6$ and SNRs $5$, $20$ and $35$ dB, is shown in Figure \ref{Fig_Convergence}.
It is evident that the algorithm eventually converges to a limit point regardless of the initialization.
However, the speed of convergence is influenced by the initial state, which also determines the limit point,  as $\overline{\bm{\mathcal{A}}}$ is non-convex.
The initialization effect becomes more visible as SNR grows large.
For example, initializing the common precoder using SVD enhances the convergence at high SNR.
In the following results, (MF-SVD) is adopted as it provides good overall performance over various channel realizations and a wide range of SNRs.
%
\begin{figure}[t!]\vspace{-5mm}
\centering
\includegraphics[width = 0.45\textwidth]{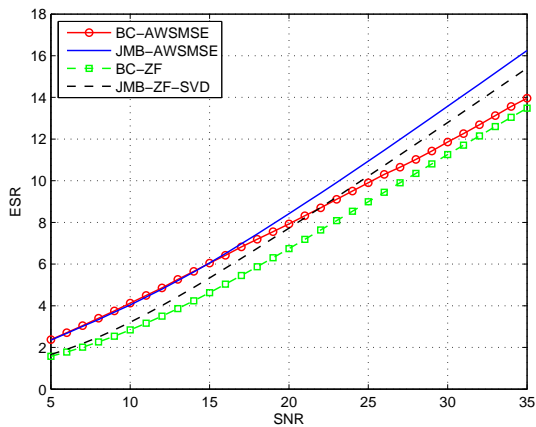}\\
\caption{MISO-BC and MISO-JMB ESRs. $K=2$, and $\alpha = 0.6$. }
\label{Fig_ESR_06}
\vspace{-5mm}
\end{figure}
%
%
\begin{figure}[t!]\vspace{-0mm}
\centering
\includegraphics[width = 0.45\textwidth]{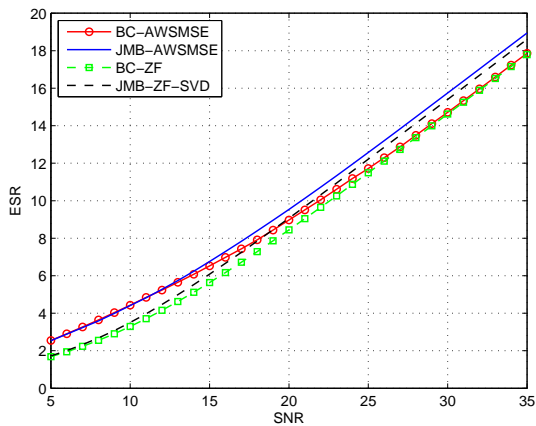}\\
\caption{MISO-BC and MISO-JMB ESRs. $K=2$, and $\alpha = 0.9$. }
\label{Fig_ESR_09}
\vspace{-5mm}
\end{figure}
%

Next, we consider the Ergodic SR (ESR) performance.
It is worth noting that the ESR is different to the ASR. The latter is the optimization metric defined in Section \ref{Subsection_MSE_MMSE_Rate}, which may not necessarily correspond to the actual SR achievable at the receivers.
The precoders obtained from optimizing the ASR yield an achievable SR defined as $R \triangleq \min_{j}\left\{ R_{\mathrm{c},j} \right\}_{j\in \mathcal{K}} + \sum_{k=1}^{K}R_{k}$, calculated using the channel realization $\mathbf{H}$.
Averaging the SR over multiple realizations of  $\mathbf{H}$ yields the ESR defined as $\mathrm{E}_{\mathbf{H}}\left\{ R \right\}$, which is used to capture the average performance over multiple channel realizations.
In the following simulations, the ESR is calculated by averaging over $100$ channel realizations.
The proposed JMB-AWSMSE scheme is compared to the conventional BC-AWSMSE scheme which corresponds to a robust adaptation of the scheme proposed in \cite{Christensen2008}.
Moreover, the base line for conventional transmission is taken as naive ZF-BF with Water-Filling (WF), i.e. optimization is carried out assuming that the estimate $\widehat{\mathbf{H}}$ is perfect, and the channel estimation error is not considered. On the other hand, we consider a modified version of the DoF-motivated scheme  in \cite{Hao2013} as a baseline for JMB.
In particular, the power splitting between the common symbols and the private symbols is maintained, while WF is used to allocate the power among the private symbols. Furthermore, the common precoder is obtained using SVD.
The ESRs for $\alpha=0.6$, and $\alpha=0.9$ are shown in Figure \ref{Fig_ESR_06} and Figure \ref{Fig_ESR_09}, respectively.
The superiority of all schemes over ZF-BF for the entire SNR range is evident.
Moreover, JMB-ZF-SVD and JMB-AWSMSE achieve the same sum DoF (slope of the curve at high SNRs).
However, the latter performs better from a SR perspective.
JMB-AWSMSE and BC-AWSMSE perform similarly at low SNRs, where the JMB's common symbol is switched off. The benefit of transmitting a common symbol manifests as SNR grows large with a gain exceeding 4 dB for $\alpha = 0.6$, in addition to the DoF gain yielding a faster increase-rate.
For $\alpha = 0.9$ which is almost ideal from a DoF perspective, the common symbol is not as instrumental as it is for lower CSIT qualities. However, ESR and DoF gains can still be observed at high SNRs.
%
\section{Conclusion}
\label{Section_conclusion}
In this paper, we addressed the problem of ASR maximization in a MISO-JMB system with partial CSIT and perfect CSIR.
The ASR problem was transformed into an augmented AWSMSE problem. The AWSMSE problem was solved using an AO algorithm which was shown to converge to a stationary point of the ASR problem.
Numerical simulations were employed to demonstrated the benefits of transmitting a common symbol in addition to the private symbols.
In particular, the rate performances of the proposed JMB scheme and a state-of-the art linearly precoded MU-MISO scheme were compared.
At high SNRs, it was shown that the gains anticipated by the DoF-based analysis are achieved with an enhanced rate performance compared to the DoF-motivated design.
On the other hand, the proposed scheme converges to conventional MU transmission whenever the common symbol is not needed, e.g. in the low SNR regime.
%
%
%
\ifCLASSOPTIONcaptionsoff
  \newpage
\fi
\bibliographystyle{IEEEtran}
\bibliography{IEEEabrv,References}

\begin{thebibliography}{10}
\providecommand{\url}[1]{#1}
\csname url@samestyle\endcsname
\providecommand{\newblock}{\relax}
\providecommand{\bibinfo}[2]{#2}
\providecommand{\BIBentrySTDinterwordspacing}{\spaceskip=0pt\relax}
\providecommand{\BIBentryALTinterwordstretchfactor}{4}
\providecommand{\BIBentryALTinterwordspacing}{\spaceskip=\fontdimen2\font plus
\BIBentryALTinterwordstretchfactor\fontdimen3\font minus
  \fontdimen4\font\relax}
\providecommand{\BIBforeignlanguage}[2]{{%
\expandafter\ifx\csname l@#1\endcsname\relax
\typeout{** WARNING: IEEEtran.bst: No hyphenation pattern has been}%
\typeout{** loaded for the language `#1'. Using the pattern for}%
\typeout{** the default language instead.}%
\else
\language=\csname l@#1\endcsname
\fi
#2}}
\providecommand{\BIBdecl}{\relax}
\BIBdecl

\bibitem{Clerckx2013}
B.~Clerckx and C.~Oestges, \emph{MIMO Wireless Networks: Channels, Techniques
  and Standards for Multi-antenna, Multi-user and Multi-cell Systems}.\hskip
  1em plus 0.5em minus 0.4em\relax Academic Press, 2013.

\bibitem{Yang2013}
S.~Yang, M.~Kobayashi, D.~Gesbert, and X.~Yi, ``{Degrees of freedom of time
  correlated MISO broadcast channel with delayed CSIT},'' \emph{IEEE
  Transactions on Information Theory}, vol.~59, no.~1, pp. 315--328, 2013.

\bibitem{Hao2013}
C.~Hao and B.~Clerckx, ``{MISO Broadcast Channel with imperfect and (Un)matched
  CSIT in the frequency domain: DoF region and transmission strategies},'' in
  \emph{IEEE 24th International Symposium on Personal Indoor and Mobile Radio
  Communications (PIMRC)}, Sept 2013, pp. 1--6.

\bibitem{Jindal2006a}
N.~Jindal and Z.-Q. Luo, ``{Capacity Limits of Multiple Antenna Multicast},''
  in \emph{IEEE International Symposium on Information Theory}, 2006, pp.
  1841--1845.

\bibitem{Viswanath2003}
P.~Viswanath and D.~Tse, ``{Sum capacity of the vector Gaussian broadcast
  channel and uplink-downlink duality},'' \emph{IEEE Transactions on
  Information Theory}, vol.~49, no.~8, pp. 1912--1921, Aug 2003.

\bibitem{Christensen2008}
S.~Christensen, R.~Agarwal, E.~Carvalho, and J.~Cioffi, ``{Weighted sum-rate
  maximization using weighted MMSE for MIMO-BC beamforming design},''
  \emph{IEEE Transactions on Wireless Communications}, vol.~7, no.~12, pp.
  4792--4799, December 2008.

\bibitem{Sidiropoulos2006}
N.~Sidiropoulos, T.~Davidson, and Z.-Q. Luo, ``Transmit beamforming for
  physical-layer multicasting,'' \emph{IEEE Transactions on Signal Processing},
  vol.~54, no.~6, pp. 2239--2251, June 2006.

\bibitem{Shi2011}
Q.~Shi, M.~Razaviyayn, Z.-Q. Luo, and C.~He, ``{An Iteratively Weighted MMSE
  Approach to Distributed Sum-Utility Maximization for a MIMO Interfering
  Broadcast Channel},'' \emph{IEEE Transactions on Signal Processing}, vol.~59,
  no.~9, pp. 4331--4340, Sept 2011.

\bibitem{Bashar2014}
M.~Bashar, Y.~Lejosne, D.~Slock, and Y.~Yuan-Wu, ``{MIMO broadcast channels
  with Gaussian CSIT and application to location based CSIT},'' in
  \emph{Information Theory and Applications Workshop (ITA)}, Feb 2014, pp.
  1--7.

\bibitem{Razaviyayn2013a}
M.~Razaviyayn, M.~Boroujeni, and Z.-Q. Luo, ``{A stochastic weighted MMSE
  approach to sum rate maximization for a MIMO interference channel},'' in
  \emph{IEEE 14th Workshop on Signal Processing Advances in Wireless
  Communications (SPAWC)}, June 2013, pp. 325--329.

\bibitem{Razaviyayn2013b}
M.~Razaviyayn, M.~Hong, and Z.-Q. Luo, ``{Linear transceiver design for a MIMO
  interfering broadcast channel achieving max–min fairness},'' \emph{Signal
  Processing}, vol.~93, no.~12, pp. 3327 -- 3340, 2013.

\bibitem{Joudeh2014a}
H.~Joudeh and B.~Clerckx, ``{Sum Rate Maximization for Linearly Precoded
  Multiuser MISO Systems with Partial CSIT: A Joint Multicasting and
  Broadcasting Approach},'' \emph{submitted to IEEE Transactions on Signal
  Processing}, 2014.

\bibitem{Grant2008}
M.~Grant, S.~Boyd, and Y.~Ye, ``{CVX: MATLAB software for disciplined convex
  programming [Online]},'' \emph{Available: http://www.stanford.edu/~boyd/cvx},
  2008.

\end{thebibliography}
\end{document}